\documentstyle[sprocl,epsf]{article}

\def\Journal#1#2#3#4{{#1} {\bf #2}, #3 (#4)}

\def\NPB{{\em Nucl. Phys.} B}
\def\PLB{{\em Phys. Lett.}  B}
\def\PRL{\em Phys. Rev. Lett.}
\def\PRD{{\em Phys. Rev.} D}
\def\ZPC{{\em Z. Phys.} C}

\def\be{\begin{equation}}
\def\ee{\end{equation}}
\def\bea{\begin{eqnarray}}
\def\eea{\end{eqnarray}}

\begin{document}

\title{Nonperturbative Contributions to the Hot Electroweak
Potential}

\author{S. J. Huber, M. G. Schmidt\footnote{Talk
presented by M. G. Schmidt at the ``Strong and Electroweak
Matter'' Conference, Copenhagen, Dec. 1998.}
}

\address{Institut f\"ur Theoretische Physik der Universit\"at Heidelberg,
\\ Philosophenweg 16, D-69120 Heidelberg, Germany
\\E-mail: s.huber@thphys.uni-heidelberg.de, m.g.schmidt@
thphys.uni-heidelberg.de}


\maketitle\abstracts{ The hot electroweak potential
for small Higgs field values is argued
to obtain contributions
from a fluctuating gauge field background leading to
confinement. The destabilization of $F^2=0$
and the crossover are discussed in our phenomenological
approach, also based on lattice data.}

\section{Introduction}

The infrared behavior of the electroweak standard model (SM)
at high temperatures $T$ is isolated most elegantly
in a truncated effective 3-dimensional action obtained
by matching it to the original theory \cite{1} in (two-loop)
perturbation theory integrating out nonzero modes and
longitudinal gauge bosons. With high accuracy
it leads to a Lagrangian with $T$-dependent couplings
\be\label{1}
L_3^{eff}=\frac{1}{4}F_{ij}^aF_{ij}^a+(D_i\phi)^+D_i\phi+
m^2_3(T)\phi^+\phi+\lambda_3(T)(\phi^+\phi)^2,\ee
where the gauge coupling $g^2_3=g_w^2T(1+...)$ sets the scale and the
dimensionful couplings $\lambda_3, m^2_3$ can be scaled
to it introducing $x=\lambda_3/g_3^2(\sim
\lambda_4/g_w^2)$, which depends on the Higgs mass and
determines the strength of the phase transition (PT),
and $y=m^2_3/g^4_3$ which is $\sim (T-T_c)$ near the
critical temperature $T_c$. (\ref{1}) describes
a superrenormalizable theory; $g_3^2$ has only infrared
running in the Wilsonian sense, its divergence in the IR signalizes
confinement \cite{2}.

Lattice calculations \cite{3,4} based on (\ref{1}) show a weakly
first-order PT for $x\leq0.11(m_H\leq0.9 m_W)$ with a crossover
above that value (predicted in previous theoretical work \cite{5}).
3-dimensional perturbation theory 
for the effective potential
always predicts a first-order
PT, mainly due to the $-(\varphi^2)^{3/2}$ term obtained from
the simple gauge boson loop in a Higgs background $\varphi^2=2\phi^+
\phi$. It strongly deviates from lattice results for $x\geq0.05$
in particular for the interface tension of the critical bubble
\cite{6} and the latent heat.

Indeed, in lattice calculations \cite{4,7} one observes typical
confinement phenomena in the hot phase (near $\varphi=0$):

i) There is a string tension $\sigma_{\rm fund}\sim 0.13\ g^4_3$ like
in pure Yang-Mills theory.

ii) There is a rich spectrum of $0^{++}("H"), 1^{--}("W"),
2^{++}$ correlation masses (see also the model in ref. 8)
including $W$-(glue)ball states  -- which do not seem to mix
with the Higgs bound states.

The previous points indicate a pure Yang-Mills dynamics with the
Higgses just sitting at the ends of the confining string. The
screening observed in ref. 9 only appears at rather large
distances and is hardly observable in the static potential.

It would be very useful to have a (coarse grained) effective
potential \linebreak 
$V_{\rm eff}(\varphi^2)$ \cite{10}, but for a gauge theory
this is a difficult task. Such a potential would help to discuss
critical bubble shapes, sphalerons, etc. A (semi)analytic model would
also allow to develop simple criteria whether one can trust
perturbative results in ``beyond'' models like the supersymmetric
models MSSM and  NMSSM, where lattice calculations become increasingly
more complicated. Here we present a phenomenological proposal
\cite{11} how to obtain nonperturbative contributions to the
effective potential of the 3-dimensional theory (\ref{1}).

\section{A Model for the Nonperturbative Part of the Effective
Potential}

In view of the Yang-Mills field dynamics argued for in the
introduction it is very natural for a small classical Higgs 
background  $\varphi$ to introduce also a gauge field background
of a 3-dimensional QCD-type vacuum \cite{12}. We want to describe the  
confinement effects mentioned above, and thus a constant field
strength background will not do. Thus we postulate a
background with Gaussian correlations, i.e. the cumulant
expansion e.g. of the Wegner-Wilson loop exponential in this vacuum
should contain only 2-correlators
\be\label{2}
\ll \exp (ig_3\oint Adx\gg\sim\exp(-\frac{1}{2}g^2_3
\oint\oint\ll AA'\gg).\ee
Using the nonabelian Stokes theorem (or the coordinate gauge
as below) this can be transferred into $\exp(-\frac{1}{2}g^2_3
\int da\int da'\ll FF'\gg$. The truncation of higher
correlators and an ansatz for the correlator $\ll FF'\gg$
defines the ``stochastic vacuum'' model \cite{13} in QCD$_4$. Indeed,
to make the correlator \linebreak
$\ll F(x)F(x')\gg$ gauge invariant,
one has to connect $x,x'$ to some reference point $x_0$ and introduce
$F(x,x_0)=P\exp(ig_3\int^x_{x_0}Adx)F(x_0)$. In the coordinate
gauge $(x-x_0)_\mu A^a_\mu=0$ and with straight line integrals
from $x_0$ to $x$ the connection vanishes and (summing over
indices) one has
\be\label{3}
\ll g^2_3F^a_{\mu\nu}(x')F^a_{\mu\nu}(x)\gg=
<g^2_3F^2>D_{\rm eff}\left(\frac{(x-x')^2}{a^2}\right).\ee
Both, $x_0$ independence and the Gaussian approximation 
are only reasonable for a choice of $x_0$  inside the loop considered.
$<g^2_3F^2>$ is the usual condensate obtained here in the limit
$a\to\infty$. A reasonable ansatz for the form factor $D$ is
$D(z^2/a^2)=e^{-|z|/a}$ with a
correlation length $a$. The latter, like in QCD$_4$,
has been determined on the lattice
for QCD$_3$ as $a^{-1}\sim 0.72\ g^2_3$.

We now want to write a nonperturbative effective potential
$V_{\rm eff}(m^2=\frac{1}{4}g_3^2\varphi^2,<g^2_3F^2>)$ in the combined Higgs
and fluctuating gauge field background. In 1-loop order this
corresponds to the graphs of type fig. 1a.

\section{Instability at $\mathbf{F^2=0}$}

\begin{figure}[t] 
\begin{picture}(200,110)
\put(140,105){\small Figure 1:}
\put(-180,-430){\epsfxsize15cm \epsffile{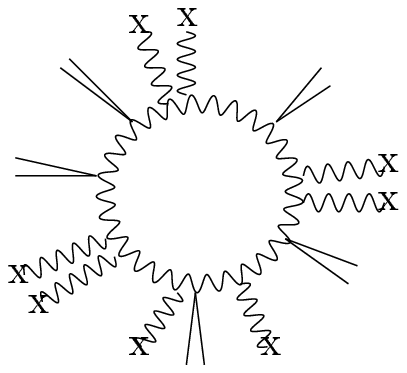}}
\put(90,-460){\epsfxsize15cm \epsffile{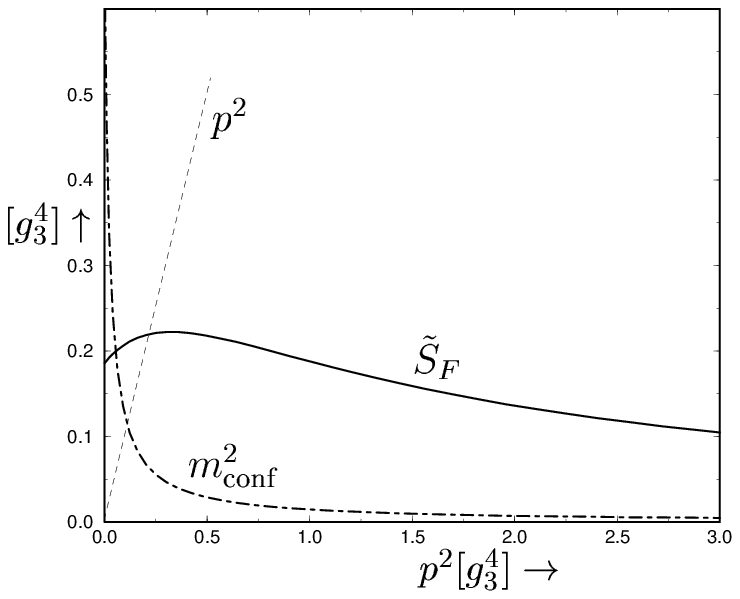}}
\put(10,90){(a)}
\put(150,90){(b)}
\end{picture} 
\end{figure}

As a first step we calculate the order $F^2$ contribution to the
effective action obtained by averaging 
the one-loop action including
ghosts (in the Feynman background gauge)
\be\label{4}
\Gamma^{\rm 1-loop}=\frac{1}{2}{\rm tr} \log[(-D^2
\delta_{\mu\nu}+2ig_3 F_{\mu\nu}+m^2\delta_{\mu\nu})^{ab}]-{\rm tr}
\log[-D^2+m^2]\ee
with $m^2=\frac{1}{4}g^2_3\varphi^2$ and a gauge boson spin interaction
term $2ig_3F_{\mu\nu}$,
over the stochastic field (according to (\ref{3})).
We obtain
\bea\label{5}
V_{FF}^{\rm 1-loop}(m^2)&=&-<g^2_3F^2>aG(ma)\nonumber\\
&=&-\frac{1}{8\pi^3}<g^2_3F^2>\int^\infty_0dp p^2
\tilde D_{\rm eff}(p^2)\int^1_0d\alpha\\
&&\{[m^2+\alpha(1-\alpha)p^2]^{-1/2}-\frac{1}{2p^2}
[(m^2+\alpha(1-\alpha)p^2)^{1/2}-m]\}\nonumber.\eea
There is a negative first term due to the spin interaction.
$V_{FF}^{\rm 1-loop}$ dominates the tree term and destabilizes $F^2=0$ if
$<\frac{1}{4}F^2>-<g^2_3F^2>aG(ma)=V_{FF}(m^2)$ is negative.
For $\varphi^2=0$ this is fulfilled for $1/a\stackrel{\scriptstyle<}
{\sim}0.6 g^2_3$ not so far from the lattice value $1/a\sim
0.72 g^2_3$. 
(In the numerical evaluation of $V_{\rm eff}$ discussed below
we took only the first term of $G(ma)$ and obtained 
instability of $F^2$$=$$0$ also for $a^{-1}\stackrel{\scriptstyle<}
{\sim}0.7 g^2_3$.)
 Lattice evaluation \cite{14} of the free energy indicates
 that $V(m^2=0,F^2_{min})$ is very small. The
negative $F^2$ coefficient then indeed would  be small and we
can hope that 2-loop contributions (or an
IR renormalization-group procedure)
lead to even better agreement.

\section{The Nonperturbative Potential}
Evaluation of the full 1-loop potential \cite{11} in the
nonperturbative gauge field background of the stochastic
vacuum in the coordinate gauge and with Schwinger
proper time/worldline methods starts from
\bea\label{6}
V(m^2,<g^2_3F^2>)&=&-\frac{1}{2}\int^\infty_0\frac{dT}{T}
\int [Dy]\exp\{-\int^T_0d\tau(\dot x^2/4+m^2)\}\\
&\times& \ll tr_{C,L}P\exp\{+ig_3\int^T_0d\tau({\bf A}_\mu\dot x_\mu
{\bf 1}_L+2{\bf F}(x)\}\gg.\nonumber\eea
Substituting ${\bf A}_\mu=\int^1_0d\eta\ \eta
y_\nu{\bf F}_{\mu\nu}
(x_0+\eta y)$, truncating the cumulant expansion and postulating
the stochastic vacuum, the correlator (\ref{3}) 
in the double area integral of the spinless minimal couplings leads
to an area law for loops of size bigger than $a$. The spin-spin
correlator comes with a destabilizing minus sign as
discussed in the previous section. We first deal with both cases
separately (and suppress the discussion of the
interference term).

a) Only spin (paramagnetic) interactions

We approximate this case by a nearest neighbor interaction and
consider the vacuum polarization $\Sigma(x,x')^{ab}_{\mu\nu}
=\delta_{\mu\nu}\delta^{ab}S_F((x-x')^2)$ of
a gauge boson  interacting with the correlated background
\be\label{7}
\tilde S_F(p^2)=<g^2F^2>\frac{1}{9\pi^2}\int^\infty_0
dq\frac{q}{p}\log((p+q)^2+m^2)/((p-q)^2+m^2))
\tilde D(q^2)\ee
which enters the log $((p^2+m^2)-\tilde S_F(p^2))$ of
the effective 1-loop potential.
It would produce an infrared instability
for small $m^2$ without further contributions.

b) Pure area law

In the model  there is obviously a relation
between the (rescaled) string tension $\bar\sigma$ appearing
in the area law $\exp(-\bar\sigma\bar A)$ and the
vacuum condensate
\be\label{8}
\bar \sigma=\frac{\pi}{9}<g^2_3F^2>T \int^\infty_0dzzD(z^2/a^2)=
\sigma_{\rm adj}T=\frac{8}{3}\sigma_{\rm fund}T.\ee
Substituting in lack of direct evaluation of the path
integral an area law in (\ref{6}) modified by some ansatz
at small areas  (see Eq. (3.5) in ref. \cite{11} for further details)
we get the renormalized expression
\bea\label{9}
V^{\rm area}&=&-\frac{3}{2}\int^\infty_0\frac{dT}{T}T^{-3/2}
\int [D\bar y]\exp\{-\int^1_0d\bar\tau\dot{\bar y}^2/4\}\nonumber\\
&&[\exp(\frac{-\sigma\bar AT^3}{\tilde c a^2/\bar A^2+T^2}-m^2T)
-1+m^2T].\eea
With the substitution $(4\pi T)^{-3/2}=\int d^3p/(2\pi)^3\exp(-p^2T)$
we can perform the $T$ integration (numerically) and enforce
the form
\bea\label{10}
V^{\rm area}&=&-\frac{3}{2}(4\pi)^{3/2}\int d^3p/(2\pi)^3
\int [D\bar y]\exp\{-\int^1_0d\bar \tau\dot{\bar y}^2/4\}\nonumber\\
&&[\log (p^2+m^2+m^2_{\rm conf}(p^2,\bar A,m^2))-\log p^2
-m^2/p^2]\eea
defining the infrared regulator ``magnetic mass'' $m^2_{\rm conf}(p^2,
\bar A,m^2)$ which starts with a term $\sim<g^2_3F^2>$. It deviates
from $\sigma \bar A$ because of the $\tilde c$ cut-off in
(\ref{9}).
We finally can superimpose both terms a) and b), if there is
no strong overlap in $p^2$ between $m^2_{\rm conf}(p^2)$ and
$\tilde S_F(p^2)$, in a potential containing $\log(p^2+m^2+m^2
_{\rm conf}(p^2\bar A,m^2)-\tilde S_F(p^2,m^2))$ and proper
renormalization and combinatorics. Of course the IR regulator $m^2
_{\rm conf}$ should be also introduced in $\tilde S_F$, eq. (\ref{7}).

\section{Evaluation and Discussion}

The unknown quantity $<g^2_3F^2>$ can in principle be obtained by
minimizing the potential $V(m^2,<g^2_3F^2>)$ with respect to $F^2$. 
Like the correlation
length $a$ it alternatively can be derived from lattice data via its
relation (\ref{8}) to the string tension. In lack of data
for $\bar A(T,m^2,\sigma)$ resp.  $\bar A(p^2,m^2,\sigma)$ and 
the area cut-off we take $\tilde c=2$ and a
function $\bar A(p^2)$ falling from $\bar A(0)\sim 2$ to $0.5$
at $p=0.4g^2_3$. This differs from our rough evaluation in ref.~\cite{11}
and brings the $F^2$ minimum in agreement with the lattice
data at $m^2=0$. Fig. 1b then shows our functions $m^2_{\rm conf}(p^2)$
and $\tilde S_F(p^2)$ and fig. 2a the potential in $<F^2>$
for $m^2$$=$$0$ $(1/a\simeq0.7 g^2_3)$.
The complicated dependence on the scales $T$ resp. $p^2$,
$m^2$, $\sigma \sim F^2$ induced by the path integral requires
further research.
The $m^2$-dependence of the quantities is in principle fixed by 
our explicit
expressions. In fig. 2b we plot  the potential
$V_{\rm tree}+V^{\rm nonpert.}_{\rm 1-loop}$ at various values of $x$
at the critical temperature using the simpleminded tuning of parameters
of ref.~\cite{11}. We then obtain the crossover
phenomenon  at $x\sim 0.11$\footnote{The crossover point with a second-order 
PT \cite{15} can
be roughly fixed \cite{11} by the postulate that the $\varphi^2$
and $\varphi^4$ terms vanish for canonical critical behavior. For values
of $x>0.11$ our potential remains convex for all temperatures.}.
Also interface tension and latent heat go to zero there.
Note again that $\varphi$ is introduced as a classical background.
We expect increasing fluctuations $<(\delta\varphi)^2>$ going to 
the crossover.

In the case of the MSSM with a ``light'' stop (talk of
Quir\'os at this conference) there is a strongly first-order PT
even at a Higgs mass $\sim$ 100 GeV, and perturbative \cite{16,18}
and lattice \cite{19} results qualitatively agree. Indeed \cite{17}
the additional graphs compared to the SM contain right-handed stops
and QCD
gluons which both do not feel $SU(2)$ nonperturbative effects.
Hence this part is well described by perturbation theory and the
standard part now is a at small effective $x$ where again perturbative
results can be trusted. This comes out more quantitatively
in our model. 
We can understand the sign of deviations:
the PT is somewhat stronger for the lattice data than for
the perturbative prediction.
In our approach this is a direct consequence of the 
$F^2$$=$$0$
instability, which lowers the potential for small values of the
Higgs field and decreases the critical temperature.
Such considerations may be particularly
valuable in cases which are not easily accessible to lattice
calculations.

\begin{figure}[t] 
\begin{picture}(200,115)
\put(152,117){\small Figure 2:}
\put(-70,-458){\epsfxsize15cm \epsffile{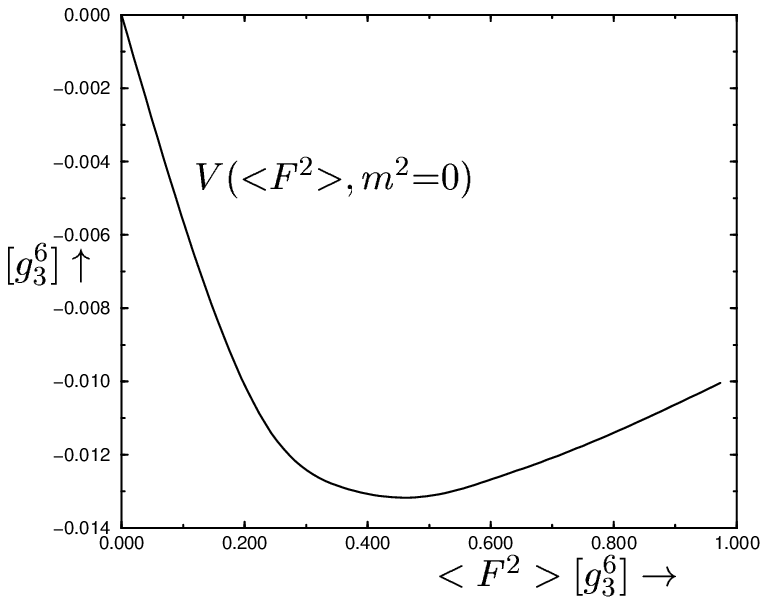}}
\put(-65,-465){\epsfxsize15cm \epsffile{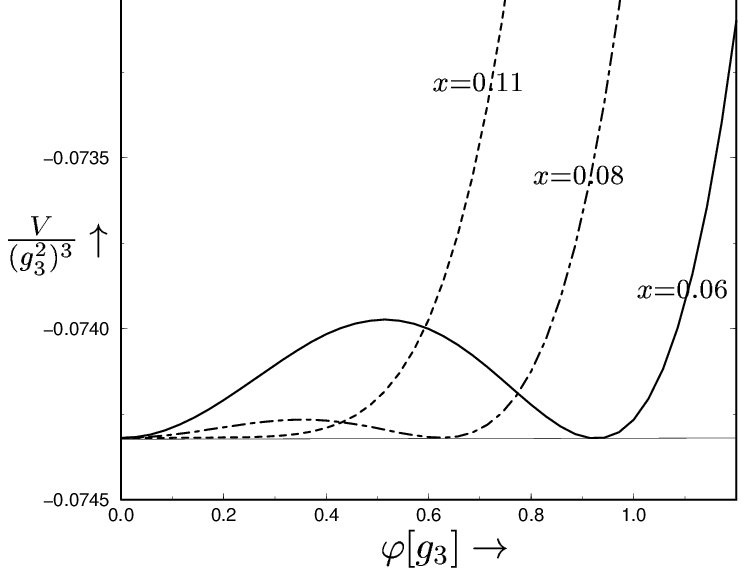}}
\put(0,95){(a)}
\put(180,95){(b)}
\end{picture} 
\end{figure}

\end{document}